# Dielectric function of aggregates of small metallic particles embedded in host insulating matrix

Leonid G. Grechko and Vitaly N. Pustovit[a]
*Institute of Surface Chemistry, NAS of Ukraine, prospekt Nauki 31, 252022 Kiev, Ukraine*

Keith W. Whites
*Department of Electrical Engineering, University of Kentucky, Lexington, Kentucky 40506-0046*

The optical properties of clusters with metallic spherical particles embedded in an insulating matrix are studied. A theoretical approach is proposed for the calculation of the macroscopic dielectric response for a collection of spheres at random positions embedded in a homogeneous medium. While accounting for the dipole–dipole interaction between particles, we have considered the frequency dependence behavior of the imaginary part of the effective dielectric constant in this system with two kinds of particles of different sizes. © *2000 American Institute of Physics.* [S0003-6951(00)05014-2]

The Maxwell-Garnett (MG)[1–3] approximation and its various updatings have frequently been used[4–13] as the description for the optical properties of composites consisting of a dielectric matrix with embedded metallic inclusions. The main purpose of[4–13] consists of accounting for the multipole interaction between inclusions. We shall consider here the method of cluster expansion developed in works.[8–11] In this method the effective dielectric constant $\tilde{\varepsilon}$ of the composite is represented as a series where each term consistently takes into account the two-particle, three-particle, and higher interactions between inclusions. The difficulty of this method is that at each stage it is necessary to solve the electrostatic problem for sets of particles (two, three, etc.) in an external field and to know the appropriate multiparticle distribution functions of inclusions in the matrix.

In the present letter, we propose to generalize the method[8] for the case of a composite containing spherical inclusions of various radii. We will take into account only the pair multipole interaction between inclusions (first correction in the MG approximation). A general expression for $\tilde{\varepsilon}$ and its imaginary part as functions of the composite parameters containing inclusions of two different radii of the same material is considered using this approximation. We shall consider a system that consists of the continuous dielectric matrix with embedded spherical particles of different kinds (noted below by indices $a,b,c...$). The dielectric permittivity of the matrix is $\varepsilon_0$ while the dielectric constants of the particles $\varepsilon_a, \varepsilon_b, \varepsilon_c...$. Let the number of spheres of kinds $a, b, c$, etc. be $N_a, N_b, N_c...$, respectively. The total number of particles is $N = \Sigma_\alpha N_\alpha$. The system is located in the external field proportional to $e^{-i\omega t}$ with a wavelength $\lambda = 2\pi c/\omega$, which is much larger than the sphere radius and mean distance between particles; $n_a = N_a/V$, $n_b = N_b/V,...$ are the concentrations of particles of the kinds $a,b,...$.

Generalizing the method of cluster expansion[8–11] in this case, we can develop the following expression for the calculation of the system effective dielectric permittivity:[14]

$$\frac{\tilde{\varepsilon}+2\varepsilon_0}{\tilde{\varepsilon}-\varepsilon_0} = \frac{1}{\frac{4\pi}{3}\Sigma_a n_a \alpha_a} - \frac{\Sigma_{a,b} n_a n_b}{(\Sigma_a n_a \alpha_a)^2}$$

$$\times \int_0^\infty R_{ab}^2 \Phi(\mathbf{R}_{ab}) d\mathbf{R}_{ab} [\beta_{ab}^{\text{II}}(\mathbf{R}_{ab}) + 2\beta_{ab}^{\perp}(\mathbf{R}_{ab})], \quad (1)$$

where

$$\alpha_a = \frac{\varepsilon_a - \varepsilon_0}{\varepsilon_a + 2\varepsilon_0} R_a^3$$

is a usual dipole polarizability of the particle of kind $a$, $\Phi_{a,b}(\mathbf{R}_{ab})$ is the two-particle distribution function of particles in the matrix, and $\mathbf{R}_{ab} = |\mathbf{r}_a - \mathbf{r}_b|$, where $\mathbf{r}_a$ and $\mathbf{r}_b$ are the origins of spheres $a$ and $b$, respectively. Equation (1) is a generalization of the relation (5.8)[12] for the case of a system with inclusions of different kinds. Taking into account only the pair dipole–dipole interaction between particles, we have[14–16]

$$\beta_{ab}^{\text{II}} = X_{10}^{(a)}(\mathbf{R}_{ab}) - \alpha_a - 2\frac{\alpha_a \alpha_b}{R_{ab}^3},$$

$$\beta_{ab}^{\perp} = X_{11}^{(a)}(\mathbf{R}_{ab}) - \alpha_a + \frac{\alpha_a \alpha_b}{R_{ab}^3}, \quad (2)$$

where $R_a$ is a radius of the particle $a$.

The coefficients $X_{10}^{(a)}(\mathbf{R}_{ab})$ and $X_{11}^{(a)}(\mathbf{R}_{ab})$ can be obtained from the solution of the problem of the electrostatic response for spheres $a$ and $b$ in the field $\mathbf{E}_0$. It should be noted that the method developed supposes generalization on the case of the higher pair multipole interactions[17] as well as on the case of multiparticle interactions. Convergence of the integral in Eq. (1) in the limit $N \to \infty$, $V \to \infty$, $N/V = $ const was discussed in detail elsewhere.[8–11] From the results of Ref. 14 it follows that

$$X_{10}^{(a)}(\mathbf{R}_{ab})$$

$$=\frac{1}{2}R_a^3\left[\frac{\left(\frac{B_a}{B_b}\right)^{1/2}+\Delta_{ab}^{3/2}}{\sqrt{\frac{1}{B_aB_b}-2\Delta_{ab}^{3/2}o_a^3}}+\frac{\left(\frac{B_a}{B_b}\right)^{1/2}-\Delta_{ab}^{3/2}}{\sqrt{\frac{1}{B_aB_b}+2\Delta_{ab}^{3/2}o_a^3}}\right], \quad (3)$$

$$X_{11}^{(a)}(\mathbf{R}_{ab})$$

$$=\frac{1}{2}R_a^3\left[\frac{\left(\frac{B_a}{B_b}\right)^{1/2}-\Delta_{ab}^{3/2}}{\sqrt{\frac{1}{B_aB_b}-\Delta_{ab}^{3/2}o_a^3}}+\frac{\left(\frac{B_a}{B_b}\right)^{1/2}+\Delta_{ab}^{3/2}}{\sqrt{\frac{1}{B_aB_b}+\Delta_{ab}^{3/2}o_a^3}}\right],$$

where $\Delta_{ab}=R_b/R_a$, $o_a=R_a/R_{ab}$, and $B_a=(\varepsilon_a-\varepsilon_0)/(\varepsilon_a+2\varepsilon_0)$; $R_a$ and $R_b$ are the radii of particles $a$ and $b$, respectively. Using the elementary approximation for the two-particle distribution function $\Phi_{ab}(\mathbf{R}_{ab})$

$$\Phi(\mathbf{R}_{ab})=\begin{cases} 1 & R_{ab}>R_a+R_b \\ 0 & R_{ab}<R_a+R_b \end{cases}, \quad (4)$$

and Eq. (1) we have

$$\frac{\tilde{\varepsilon}+2\varepsilon_0}{\tilde{\varepsilon}-\varepsilon_0}=\frac{1}{\frac{4\pi}{3}\sum_a n_a B_a R_a^3}-\frac{\Sigma_{a,b}n_an_bB_aB_bR_a^3R_b^3}{3(\Sigma_a n_aB_aR_a^3)^2}\left\{\left[1+\left(\frac{R_a^3}{R_b^3}\right)^{1/2}\left(\frac{B_b}{B_a}\right)^{1/2}\right]\ln\frac{(R_a+R_b)^3+(R_a^3R_b^3B_aB_b)^{1/2}}{(R_a+R_b)^3-2(R_a^3R_b^3B_aB_b)^{1/2}}\right.$$
$$\left.+\left[1-\left(\frac{R_a^3}{R_b^3}\right)^{1/2}\left(\frac{B_b}{B_a}\right)^{1/2}\right]\ln\frac{(R_a+R_b)^3-(R_a^3R_b^3B_aB_b)^{1/2}}{(R_a+R_b)^3+2(R_a^3R_b^3B_aB_b)^{1/2}}\right\}. \quad (5)$$

---

[a]Author to whom correspondence should be addressed; electronic mail: Pustovit.Vitaly@angstrom.uu.se

In the case of one kind of particle, from Eq. (5) we find

$$\tilde{\varepsilon}=\varepsilon_0\left(1+\frac{3fB_a}{1-fB_a-\frac{2}{3}fB_a\ln\frac{8+B_a}{8-2B_a}}\right), \quad (6)$$

where $f=(4\pi/3)R^3n$, $B_a=(\varepsilon_a-\varepsilon_0)/(\varepsilon_a+2\varepsilon_0)$, and $n$ is the concentration of inclusions. This coincides with formula (3.9).[12]

Now we consider the case of two kinds of particles $n_a=n_b=n_0$; $B_a=B_b=B$, assuming that $R_b\neq R_a$, and $\Delta_{ab}=\Delta=(R_b/R_a)<1$. Then from Eq. (5) we find

$$\tilde{\varepsilon}=\varepsilon_0\left[1+\frac{3f_0(1+\Delta^3)}{\frac{1}{B}-f_0(1+\Delta^3)-\frac{2}{3}f_0D}\right], \quad (7)$$

$$D=\left(\frac{1+\Delta^6}{1+\Delta^3}\right)\ln\frac{8+B}{8-2B}$$
$$+\frac{\Delta^3}{2(1+\Delta^3)}\left[\left(\Delta^{3/4}+\frac{1}{\Delta^{3/4}}\right)^2\ln\frac{(1+\Delta)^3+B\Delta^{3/2}}{(1+\Delta)^3-2B\Delta^{3/2}}\right.$$
$$\left.-\left(\Delta^{3/4}-\frac{1}{\Delta^{3/4}}\right)^2\ln\frac{(1+\Delta)^3-B\Delta^{3/2}}{(1+\Delta)^3+2B\Delta^{3/2}}\right], \quad (8)$$

where $f_0=(4\pi/3)R_a^3n_0$.

Using this formula, we have carried out the numerical calculation of the frequency dependencies of the imaginary part of $\tilde{\varepsilon}$ at various parameters for a composite consisting of a glass matrix with embedded silver inclusions. The dielectric function of the matrix is presented as $\varepsilon_0=2.25$ and the dielectric function $\varepsilon(\omega)$ of the metallic spheres is given by the Drude model

$$\varepsilon(\omega)=\varepsilon_\infty'+i\varepsilon_\infty''-\frac{\omega_p^2}{\omega(\omega+i\gamma)}, \quad (9)$$

with $\varepsilon_\infty'=4.5$, $\varepsilon_\infty''=0.16$, $\omega_p=1.46\times 10^{16}\,\text{s}^{-1}$, $\gamma=1.68\times 10^{14}\,\text{s}^{-1}$ for silver spheres. The results of these calculations for the frequency dependencies of Im $\tilde{\varepsilon}(\omega)$ according to Eq. (7) at some values of the parameter $\Delta$ are presented in Fig. 1. The value of $f_0$ is chosen to be equal 0.04. Using Eq. (9) at $\varepsilon_\infty''=0$ and $\gamma=0$, we can obtain that at the frequencies

$$\omega_{(1,2)}^2=\omega_s^2\frac{1\mp 2\Delta^{3/2}o_a^3}{1\mp 2\Delta^{3/2}o_a^3B(\infty)};$$

$$\omega_{(3,4)}^2=\omega_s^2\frac{1\mp\Delta^{3/2}o_a^3}{1\mp\Delta^{3/2}o_a^3B(\infty)}; \quad (10)$$

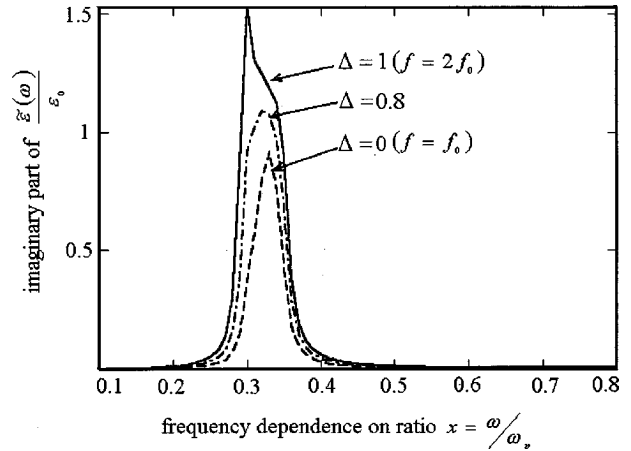

FIG. 1. Plot of the imaginary part of $\tilde{\varepsilon}(\omega)/\varepsilon_0$ depending on $x=\omega/\omega_p$ (where $\omega_p$ is a plasma frequency) and parameter $\Delta$ for silver spheres in glass at volume fraction $f_0=0.04$.

$$B(\infty) = \frac{\varepsilon'_\infty - \varepsilon_0}{\varepsilon'_\infty + 2\varepsilon_0},$$

expressions (3) have singularities [here $\omega_s^2 = \omega_p^2/(\varepsilon'_\infty + 2\varepsilon_0)$ and $o_a = R_a/R_{ab}$]. These frequencies are the surface dipole plasma modes of two particles with different radii.[18,19]

We will now briefly discuss the results obtained. Let us first consider the case when $\Delta = 1$. Then from Eqs. (7) and (8) it follows that $\tilde{\varepsilon}$ can be found from Eq. (6) at $f = 2f_0$, which reduces to the MG approximation if the logarithmic term in the denominator of Eq. (6) is neglected. This term is associated with the pair dipole–dipole interaction (PDDI) between inclusions. However, the physical picture of the interaction of the radiation with the metallic composite in these two cases differs. In the MG approximation, the absorption peak of the radiation corresponds to the frequency

$$\omega_0 = \omega_s \sqrt{\frac{1-f}{1-f\left(\frac{\varepsilon'_\infty - \varepsilon_0}{\varepsilon'_\infty + 2\varepsilon_0}\right)}},$$

which at very small $f$ coincides with the frequency of the surface plasmon $\omega_s$ of a separate inclusion.

Taking into account the PDDI results in the appearance of the limited spectrum of frequencies instead of one frequency $\omega_s$ in the system. The borders of this spectrum can be determined from consideration of the problem of two-particle interaction. Actually, at $\Delta = 1$ and $\varepsilon''_\infty \to 0$, $\gamma \to 0$ from Eq. (10) it follows that each particle can absorb at the two frequencies

$$\omega_{\text{II}}^2 = \omega_s^2 \frac{1 - 2o^3}{1 - 2o^3 B(\infty)}; \quad \omega_\perp^2 = \omega_s^2 \frac{1 + o^3}{1 + o^3 B(\infty)}, \quad (11)$$

whose values essentially depend on the distance $R$ between the given particle and any other particle of the system. [Two other frequencies in Eq. (10) we do not take into account because the numerators in Eq. (3) at $\Delta = 1$ are equal to zero.] Note that as $R \to \infty (f \to 0)$, $\omega_{\text{II}}^2 = \omega_\perp^2 = \omega_s^2$ and at $R = 2R_a$ (minimal distance between particles) these frequencies are given by the expressions

$$\bar{\omega}_{\text{II}}^2 = \frac{\omega_p^2}{\varepsilon'_\infty + 3\varepsilon_0}; \quad \bar{\omega}_\perp^2 = \frac{\omega_p^2}{\varepsilon'_\infty + \frac{5}{3}\varepsilon_0}. \quad (12)$$

All other possible frequencies determined from Eq. (10), at $2R_a < R < \infty$, will be in an interval $(\bar{\omega}_{\text{II}}, \bar{\omega}_\perp)$. A width of this interval

$$\nu^2 = \frac{4\omega_p^2}{(3\varepsilon'_\infty + 5\varepsilon_0)(\varepsilon'_\infty + 3\varepsilon_0)}$$

is maximal at $\varepsilon_\infty = \varepsilon_0 = 1$, $\nu^2 = \omega_p^2/8$. A width of the interval for frequencies $\omega > \omega_s$ is less than the width of the interval for $\omega < \omega_s$. Their ratio is equal to $(\varepsilon'_\infty + 3\varepsilon_0)/(3\varepsilon'_\infty + 5\varepsilon_0)$

$< 1$. Thus, at small $f$ (distance $R$ between particles is large), the absorption in the system occurs basically at the frequencies close to $\omega_s$. With increasing $f$ there is a displacement of the absorption frequency from frequency $\omega_s$ to the frequencies $\bar{\omega}_{\text{II}}$ or $\bar{\omega}_\perp$ depending on arrangement of pair of particles with respect to the external field $\mathbf{E}_0$. The spectral dependence in $\tilde{\varepsilon}(\omega)$ for the composite is taken after averaging for all possible pair positions of the particles in the matrix

It should be noted that in the metallic composite at $f \sim 0.1$ and more, one could observe the fine structure of the spectrum[4] if only the PDDI is taken into account. Accounting for higher pair interactions between inclusions (quadrupole with $l = l' = 2$, octupole with $l = l' = 3$, etc.) can be made within the framework of our theory and results in some partial smoothing of the frequency dependencies of Im $\tilde{\varepsilon}(\omega)$.[17] The familiar effects of smoothing, and other reasons including three-particle, four-particle, and higher interactions, could be caused by the effects of particle clustering and so on. Especially essential is the character of dependency of Im $\tilde{\varepsilon}(\omega)$ on the distribution of particle sizes. The numerical calculations of Im $\tilde{\varepsilon}(\omega)$ (Fig. 1) for the case of the system with silver particles of two different radii have shown that there is smoothing in the frequency dependence of Im $\tilde{\varepsilon}(\omega)$ and also a lowering of the peak in comparison with the case $\Delta = 1$. The last one is caused by the fact that at $\Delta \to 0$, the ratio Eq. (7), reduces to Eq. (6) with $f = f_0$, i.e., the contribution of particles with small radii to $\tilde{\varepsilon}$ becomes insignificant.

[1] C. Maxwell-Garnett, Philos. Trans. R. Soc. London, Ser. A **203**, 385 (1904).
[2] C. F. Bohren and P. R. Huffman, *Absorption and Scattering of Light by Small Particles* (Wiley, New York, 1983).
[3] U. Kreibig and M. Vollmer, *Optical Properties of Metal Clusters*, Springer Series in Material Science (Springer, Berlin, 1995), Vol. 25.
[4] W. Lamb, D. M. Wood, and N. W. Ashcroft, Phys. Rev. B **21**, 2248 (1980).
[5] R. G. Barrera, G. Monsivais, W. L. Mochan, and E. Anda, Phys. Rev. B **39**, 9998 (1989).
[6] A. Liebsch and B. N. J. Persson, J. Phys. C **16**, 5375 (1983).
[7] V. A. Davis and L. Schwartz, Phys. Rev. B **31**, 5155 (1985).
[8] V. M. Finkelberg, Sov. Phys. JETP **19**, 494 (1964).
[9] V. M. Finkelberg, Sov. Phys. Dokl. **8**, 907 (1964).
[10] B. U. Felderhof, G. W. Ford, and E. G. D. Cohen, J. Stat. Phys. **28**, 135 (1982).
[11] B. Cichocki and B. U. Felderhof, J. Stat. Phys. **53**, 499 (1988).
[12] B. U. Felderhof and R. B. Jones, Phys. Rev. B **39**, 5669 (1989).
[13] B. U. Felderhof, Physica A **207**, 13 (1994).
[14] L. G. Grechko, A. Yu. Blank, V. V. Motrich, and A. O. Pinchuk, Radiophys Radioastronomy **2**, 19 (1997). (Ukraine)
[15] B. U. Felderhof, G. W. Ford, and E. G. D. Cohen, J. Stat. Phys. **28**, 649 (1982).
[16] D. Stroud and F. P. Pan, Phys. Rev. B **17**, 1602 (1978).
[17] L. G. Grechko and V. N. Pustovit, Proceedings of Bianisotropics '97, edited by Werner S. Weiglhofer, Glasgow, 1997, p. 227.
[18] J. M. Gerardy and M. Ausloos, Phys. Rev. B **26**, 4703 (1982).
[19] M. Inoue and K. Ohtaka, J. Phys. Soc. Jpn **52**, 3853 (1983).